\documentclass[a4paper,11pt]{article}
\pdfoutput=1 

\usepackage{jinstpub} 
\usepackage[english]{babel}
\usepackage{float}
\title{\boldmath Simulation of particle identification with the cluster counting technique}

\emailAdd{federica.cuna@le.infn.it, giovanni.tassielli@le.infn.it}

\author[a,b,1]{F. Cuna \note{Corresponding author.}}
\author[c,e]{, N. De Filippis}
\author[a]{, F. Grancagnolo}
\author[c,d,1]{, G. F. Tassielli}
\affiliation[a]{Istituto Nazionale di Fisica Nucleare Sezione di Lecce, Via per Arnesano, 73100 Lecce, Italy}
\affiliation[b]{Dipartimento di Matematica e Fisica "Ennio De Giorgi", Universit\`{a} del Salento, Via per Arnesano, 73100 Lecce, Italy}
\affiliation[c]{Istituto Nazionale di Fisica Nucleare Sezione di Bari, Via E. Orabona 4, 70125 Bari, Italy}
\affiliation[d]{Universit\`{a} degli Studi di Bari "Aldo Moro", Piazza Umberto I, 1, Bari}
\affiliation[e]{ Politecnico di Bari, Via Amendola 126/b, Bari, Italy}

\abstract{\noindent 
In this paper we show the potential of the cluster counting technique for particle identification. Simulations based on \textit{Garfield++} software prove that this technique improves the particle separation capabilities with respect to the ones obtained with the traditional method of dE/dx.\\
Moreover three different algorithms to reproduce the clusters number and the cluster size distribution with \textit{Geant4} software are discussed.
}

\proceeding{Talk presented at the International Workshop on Future Linear Colliders (LCWS2021), 15-18 March 2021. C21-03-15.1.
 }

\begin{document}
\maketitle
\flushbottom

\section{Introduction}
\label{sec:intro}
\begin{figure}[H]
	\centering
	\includegraphics[width=0.5\textwidth]{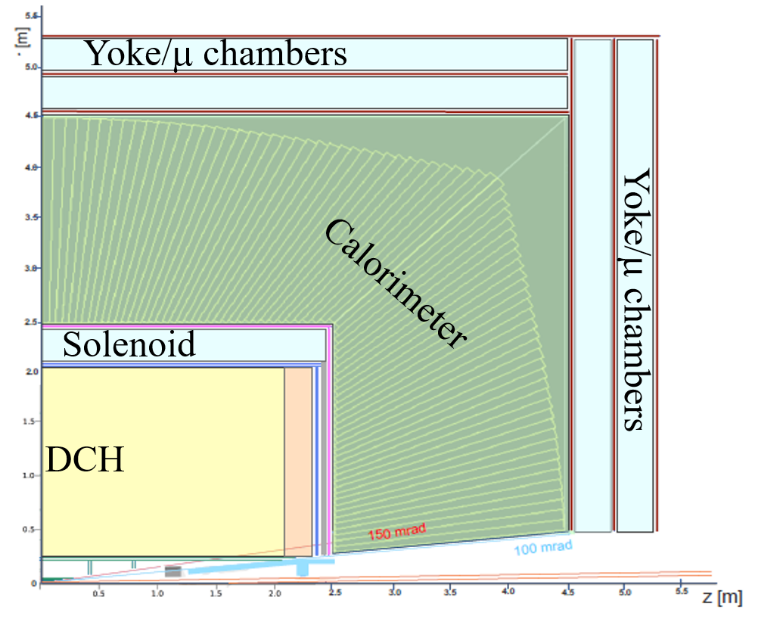}
	\caption{Schematic layout of \textbf{IDEA} detector.}
	\label{idea}
\end{figure}

Particle identification is one of the most important and difficult task for the experimental high energy physics. There are two main ways to identify particles: the first is by analysing the way in which they interact with the detector, ie. by studying the signature that they leave. As an example, if a particle is detected only in a electromagnetic calorimeter, we can conclude that it is a photon. The other way is by determining their mass and charge. In particular, the mass is derived by the simultaneous measurements of the momentum, measured thanks to the curvature of the track, in a suitable magnetic field and velocity, achievable, among the other methods, by the measurement of the energy deposit by ionization \cite{lipman}.\\
Indeed, the ionization of matter by charged particles is the primary mechanism used for particle identification (dE/dx), but the large uncertainties in the total energy deposition represent a limit to the particle separation capabilities.\\
The cluster counting technique (dN/dx) takes advantage from the primary ionization \textit{Poissonian} nature and offers a more statistically significant way to infer the mass information \cite{cc}. \\
This technique will be widely explored with the \textbf{IDEA} drift chamber \cite{idea1}. \\
\textbf{IDEA} is the new detector concept intended for both the FCC-ee collider, proposed to be built at CERN, and the CEPC collider, proposed to be built in China \cite{idea1}. The schematic layout is shown in Figure \ref{idea}.\\
Our goal is to investigate the potential of the cluster counting technique, which means by implementing reasonable and fast algorithms to reproduce the clusters number distribution and the cluster size distribution in \textit{Geant4} \cite{geant4}, a powerful software commonly used to simulate complex detector behaviour and collider events.\\
In section 2 we describe briefly the main features of the \textbf{IDEA} drift chamber. Section 3 is devoted to a description of the cluster counting technique. Section 4 deals with the algorithm development and analysis of simulation results obtained with \textit{Garfield++} software \cite{Garfield}. Section 5 describes the results obtained by implementing the algorithms with \textit{Geant4} simulation.

\section{IDEA drift chamber}
The \textbf{IDEA} drift chamber (DCH) is designed to provide an efficient tracking, a high precision
momentum measurement and an excellent particle identification by exploiting the application of the cluster counting technique.\\
The distinctive element of the chamber is its high transparency, in terms of radiation
lengths, obtained thanks to the novel approach adopted for the wiring and assembly procedures. Indeed, the total amount of material in radial direction, towards the barrel calorimeter is of the order of 1.6 $\% X_{0}$, in the forward and backward directions it is around 5.0 $\% X_{0}$, including the end-plates instrumented with front-end electronics \cite{dchidea}.\\
The feature of high transparency plays a key role for precision electroweak physics at the Z pole and for flavour physics, where the average charged particles momenta are in a range over which the multiple scattering contribution to the momentum resolution is significant.\\
The chamber is a unique volume, high granularity, all stereo, low mass cylindrical drift chamber, co-axial to the 2T solenoid field filled with helium based gas mixture. It extends from 35 cm inner radius to 2 m outer radius, for 4 m length and consists of 112 co-axial layers, at alternating sign stereo angles (in the range from 50 mrad to 250 mrad), arranged in 24 identical azimuthal sectors. The square cell size (5 field wires per sense wire) varies between 12.0 and 14.5 mm for a total of 56,448 drift cells \cite{dchidea}.\\
A system of tie-rods redirects the wire tension stress to the outer end-plate rim, where a cylindrical carbon fibre support structure, bearing the total load, is attached. Two thin carbon fibre domes enclose the gas volume. Their profile is suitably shaped in order to minimise the stress on the inner cylindrical wall and they are free to deform under gas pressure variations, without affecting the wire tension \cite{dchidea}.\\
The choice of an extremely light gas mixture allows the exploitation of the cluster counting technique that can reach a spatial resolution under 100 $\mu$m (tested in a 8 mm drift cell) and timing techniques which can reach a dN/dx resolution under 3 $\%$ for particle identification (a factor 2 better than dE/dx) \cite{dchidea}.
\section{Principles of cluster counting techniques}
\begin{figure}[H]
	\centering
	\includegraphics[width=0.2\textwidth]{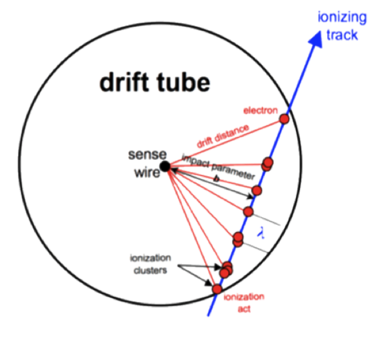}
	\includegraphics[width=0.5\textwidth]{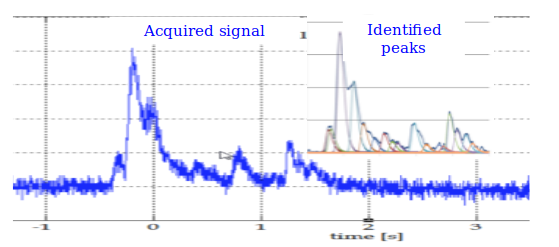}
	\caption{Left: The section of a drift tube, with an ionizing track (blu arrow) and some ionization clusters (red dots). Right: A typical signal with the identified peaks.}
	\label{drifttube}		
\end{figure}

In general, the drift chambers can provide a measurement of the energy loss along the particle trajectory, that, if associated with a measurement of the momentum, allows to derive the mass of a ionizing particle (the traditional dE/dx method).\\
However, the large and inherent uncertainties in the total energy deposition, described by Landau function, represent a serious limit to the particle identification capabilities.\\
The cluster counting technique takes advantages of \textit{Poissonian} nature of the primary ionization and offers a more statistically significant way to infer mass information. The method consists in singling out, in ever recorded detector signal, the isolated structures related to the arrival on the anode wire of the electrons belonging to a single ionization act (see Figure \ref{drifttube}).\\
\subsection{Particle identification with dE/dx and dN/dx}
\begin{figure}[H]
	\centering
	\includegraphics[width=0.5\textwidth]{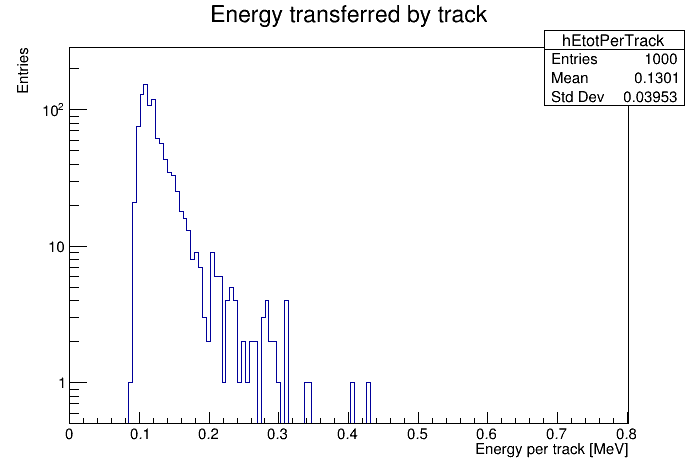}
	\caption{Energy loss distribution of a muon traversing 200 cells, 1 cm per side, filled with 90 $\%$ He and 10 $\%$ $iC_{4}H_{10}$ simulated by \textit{Garfield++}.}
	\label{Emu300}		
\end{figure}

\begin{figure}[H]
	\centering
	\includegraphics[width=0.5\textwidth]{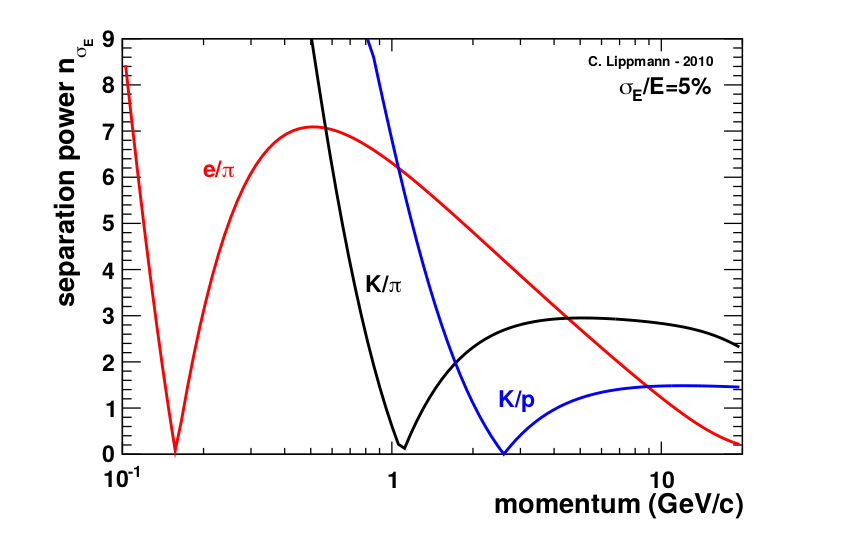}
	\caption{Typical separation power obtained with dE/dx method in a gaseous detector \cite{lipman}.}
	\label{dedx1}		
\end{figure}

\begin{figure}[h]
	\centering
	\includegraphics[width=0.5\textwidth]{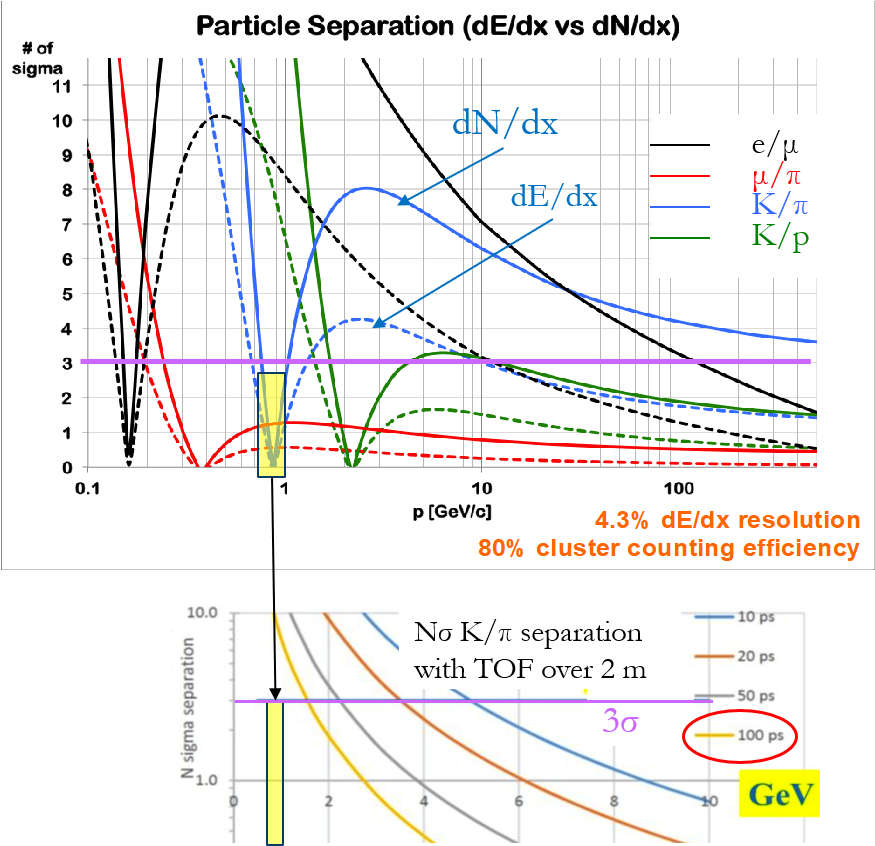}
	\caption{Top: Analytic evaluation of particle separation capabilities achievable with dE/dx (solid curves) and dN/dx (dashed curves). The region between 0.85 GeV/c and 1.05 GeV/c where a different technique is needed is highlighted in yellow. Bottom: PID performance as function of the time resolution by using a time of flight technique to recover the particle identification in the range 0.85 GeV/c and 1.05 GeV/c.}
	\label{cc1}		
\end{figure}

A particle passing through a material undergoes a series of inelastic collisions with the atomic electrons of the material. As a result, each atom could be excited or ionised, while the particle loses a small fraction of its kinetic energy. \\
Measurements of the deposited energy are widely used for particle identification. Gaseous counters (but also solid state counters) provide signals whose pulse height is proportional to the number of electrons produced in the ionization process along the track length inside the detector and thus proportional to the deposit energy.\\
The distribution of the deposit energy follows the \textit{Landau} function (see Figure \ref{Emu300}), since it allows the possibility of large energy transfers in single collisions that add a long tail (\textit{Landau} tail) to the high energy side, resulting in a asymmetric shape  whose mean value is significantly higher than the most probable value \cite{lipman}.\\
\noindent
This implies that the mean value is not a good estimator for the energy deposition and usually a truncated mean (typically from 40 $\%$ up to 80 $\%$) is used.\\
The separation power for two particles, labelled for simplicity $p_{1}$ and $p_{2}$, with different masses and same momentum, is evaluated with the relation \eqref{eqp} \cite{lipman}:
\begin{equation}
n_{\sigma_{E}}=\dfrac{\Delta_{p_{1}}-\Delta_{p_{2}}}{<\sigma_{p_{1},p_{2}}>}
\label{eqp}
\end{equation}
where $\Delta_{p_{1}}$ and $\Delta_{p_{2}}$ are the measurements of the deposited energy, $\sigma_{E}$ is the resolution in the ionization measurement (\textit{energy resolution}) given by the variance of \textit{Gaussian} distribution of the truncated mean values and $<\sigma_{p1,p2}>$ is the average of the two resolutions:
\begin{equation}
<\sigma_{p1,p2}>=\dfrac{\sigma_{E,p1}+\sigma_{E,p2}}{2}
\end{equation}
A typical separation power achievable with ionization measurements in a gaseous detector with energy resolution of 5 $\%$ is shown in Figure \ref{dedx1}.\\
\noindent
Nevertheless, we can follow also another way to perform the particle identification, by studying the clusters distribution generated by a charged particle in a gas detector.
\noindent
The process of energy loss of a charged particle crossing a medium is a discrete process: a particle traversing a gas leaves a track of ionization consisting of a sequence of clusters with one or more electrons which are all released in a single act of \textbf{primary ionization}. This is a typical \textit{Poissonian} process, i.e. a result of a number of the order of the Avogadro number of unlikely independent random events, whose sum gives the mean specific ionization \cite{cc}.\\ 
The main advantage of the \textit{Poissonian} distribution is that its \textit{Gaussian} limit is achieved when the mean value reaches 20, that is of the order of 1 cm track length for the most commonly used gas mixtures. Instead, the energy distribution follows a \textit{Gaussian} shape just in thick and dense material (because of \textit{central limit theorem}), but this situation is not helpful for the drift chambers \cite{cc}.\\
We expect that the cluster counting could improve the particle identification capabilities and the analytical results shown in Figure \ref{cc1} confirm this expectation.\\
\noindent
The plot shows the particle separation power in terms of the numbers of standard deviation (sigma) as a function of momentum in a mixture of 90 $\%$ He and 10 $\%$ $iC_{4}H_{10}$.\\
Solid curves refer to separation with the cluster counting technique and dashed curves refer to the optimal energy loss truncated mean technique. A cluster counting efficiency of 80 $\%$ is assumed in the calculations. \\ 
It is evident that a relative gain of a factor \textbf{2} of the cluster counting technique with respect to the dE/dx method is reached. The technique failed just in small gaps along the momentum range analysed. For example, in the kaon-pion separation (blue curves), the technique failed around 1 GeV, but this gap could be easily recovered with a time of flight technique with a detector having a time resolution of around 100 ps.\\
\newpage
\section{Cluster counting simulation in Garfield++}
To investigate the potential of the cluster counting techniques (for He based drift chamber) on physics events a reasonable simulation of the ionization clusters generation is needed.\\
\textit{Garfield++} and \textit{Geant4} are two valid software tools for the drift chamber simulations. \\
\textit{Garfield++} can describe in detail the properties and the performance of a single cell or drift chambers, but it is not suitable to simulate a large scale detector and to study colliders events. On the other side, \textit{Geant4} can simulate elementary particle interactions with the material of a (complex) detector and study collider events, but the fundamental properties and the performances of the sensible elements, like the drift cells, have to be parametrized or "ad-hoc" physics models have to be defined.\\
Our goal is to develop an algorithm which can use the energy deposit information provided by \textit{Geant4} to reproduce, in a fast and convenient way, the clusters number distribution and the cluster size distribution.\\
We could also create a physics model, which, once integrated in \textit{Geant4}, could reproduce in detail the ionization process, but this approach would cost a great space disk occupation and a long computational time. A simple algorithm which, by using the energy deposit simulated by the software, provides us all the necessary information seems to be a better solution.\\	
For this purpose, we simulated tracks crossing  200 cells, 1 cm per side, filled with a mixture of 90 $\%$ He and $10\%$ $iC_{4}H_{10}$.
\subsection{Garfield++ simulation analysis}
\begin{figure}[H]
	\centering
	\includegraphics[width=0.4\textwidth]{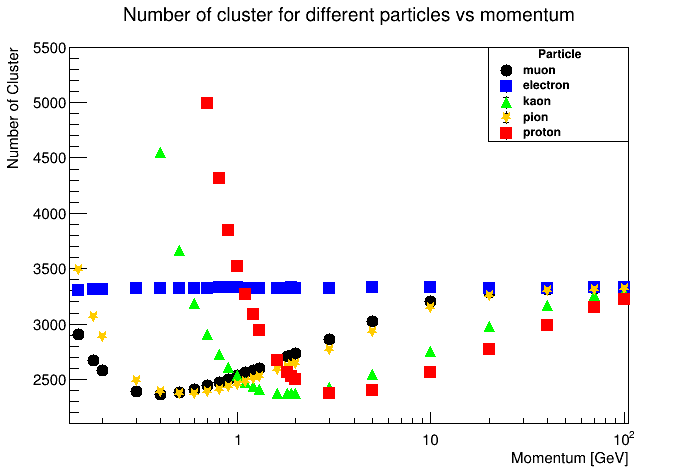}
	\includegraphics[width=0.4\textwidth]{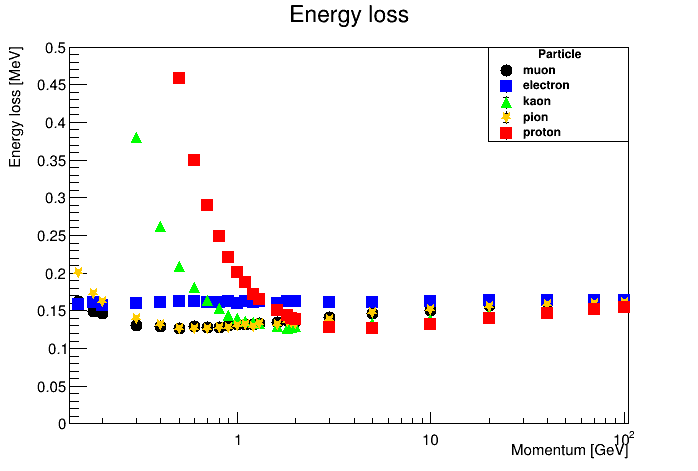}
	\caption{Left: Clusters number distribution. Right: Energy loss distribution.}
	\label{ncl}		
\end{figure}

\begin{figure}[H]
	\centering
	\includegraphics[width=0.4\textwidth]{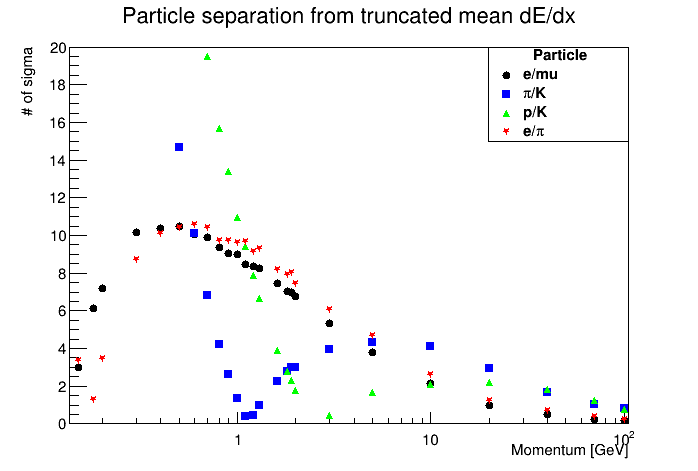}
	\includegraphics[width=0.4\textwidth]{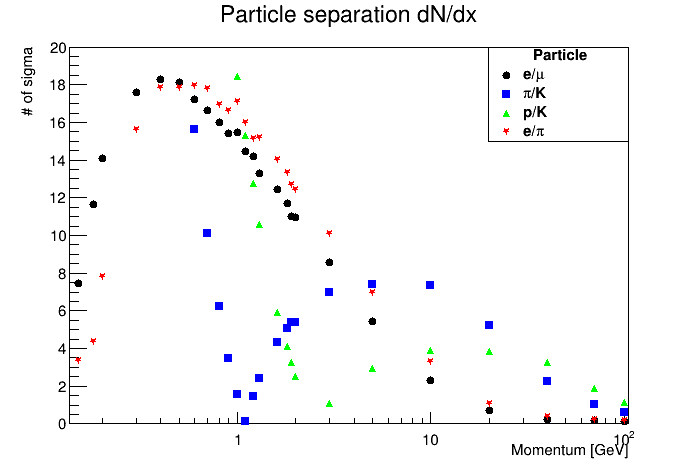}
	\caption{Left: Particle separation power with dE/dx (truncated mean at 70 $\%$). Right: Particle separation power with dN/dx.}
	\label{pow-gar}		
\end{figure}

We started studying the clusters number distribution and the energy loss distribution for muons, pions, electrons, protons and kaons in a range of momentum from 200 MeV up to 1 TeV, reported in Figure \ref{ncl}.\\
\noindent
Then we evaluated the particle separation power implementing both methods, the dE/dx one with a truncated mean of 70 $\%$ and the dN/dx one, reported in Figure \ref{pow-gar}.
\noindent
The plots show clearly that the cluster counting technique improves particle separation capabilities of a factor of \textbf{2} compared to the traditional method.\\ 
As example, around 5 GeV, the power separation of a pion from kaon (blu squared) obtained with the traditional method is around 4 sigma while the one obtained with the cluster counting technique is around 8 sigma.\\
Our goal is to obtain the same results by using \textit{Geant4}. For this purpose, we implemented three different versions of an algorithm which reconstructs the clusters number distribution and the cluster size distribution using the information given by \textit{Geant4}.\\
\subsection{Algorithm implementation}
\begin{figure}[H]
	\centering
	\includegraphics[width=0.4\textwidth]{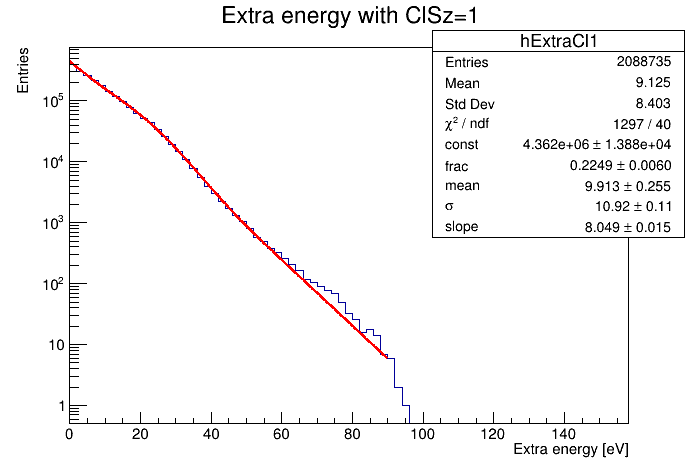}
	\centering
	\includegraphics[width=0.4\textwidth]{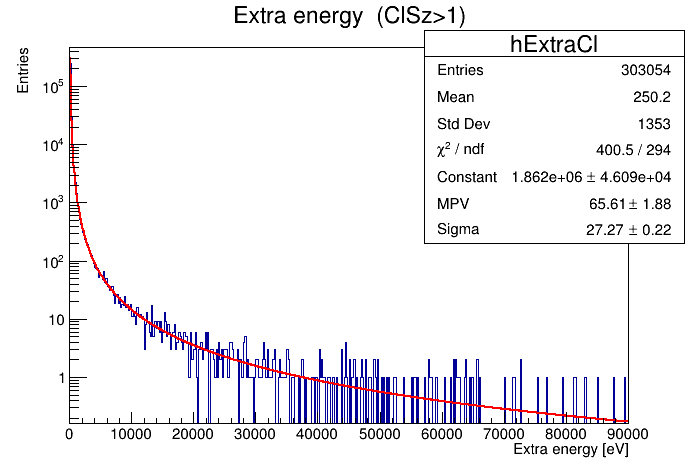}
	\caption{ Kinetic energy distribution for a muon at 300 MeV for cluster with cluster size equal to 1 (left) and for cluster size higher than 1 (right).}
	\label{cl1}		
\end{figure}

\begin{figure}[H]
	\centering
	\includegraphics[width=0.5\textwidth]{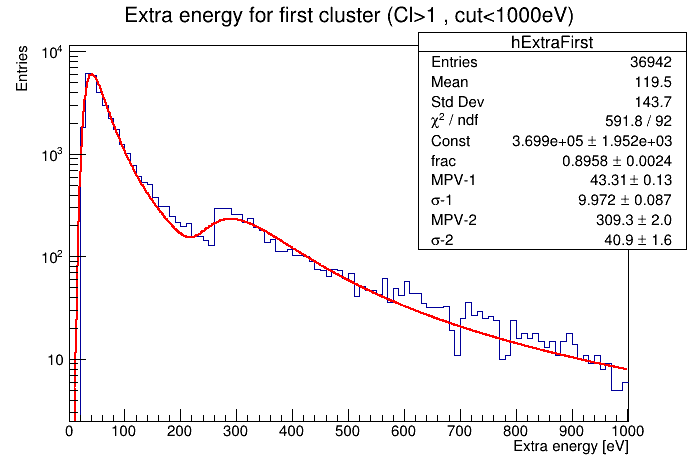}
	\caption{Kinetic energy distribution for cluster with cluster size higher than one for a muon at 300 MeV, up to 1 keV.}
	\label{cc2cut}		
\end{figure}

\begin{figure}[h]
	\centering
	\includegraphics[width=0.6\textwidth]{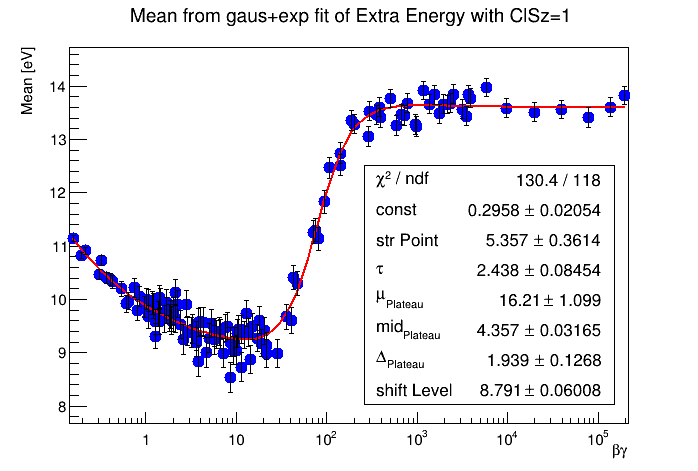}
	\caption{Mean value distribution of \textit{Gaussian} fit of the kinetic energy distribution for clusters with cluster size equal to 1, fitted with an exponential plus a plateau function.}
	\label{meanextra1}		
\end{figure}

\begin{figure}[h]
	\centering
	\includegraphics[width=0.6\textwidth]{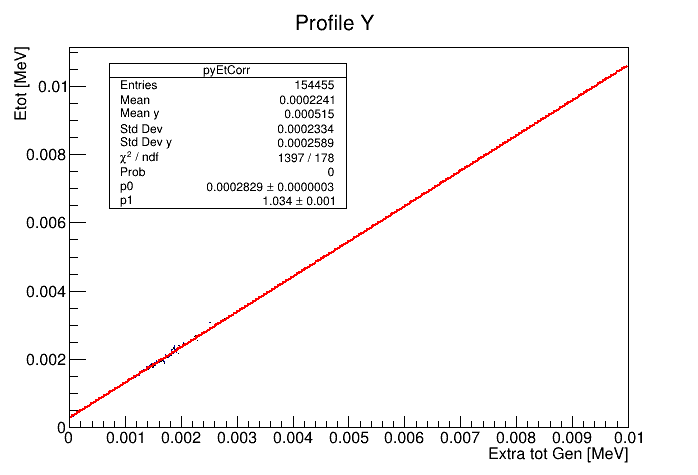}
	\caption{Correlation plot between the total energy loss  by a muon at 300 MeV traversing gas and the total kinetic energy for clusters with cluster size higher than one.}
	\label{maxExEcl}		
\end{figure}

The algorithm implementation starts from \textit{Garfield++} simulations.\\
Firstly, we analyse the distribution of the kinetic energy for clusters that have a cluster size equal to 1, as shown on the left side of Figure \ref{cl1} and clusters that have cluster size higher than 1, as shown in the right side of Figure \ref{cl1}. 
\noindent
Moreover, we study separately the distribution of cluster with a cluster size higher than 1 up to a 1 keV, which is a cut equivalent to the single interactions range cut set by default in \textit{Geant4}.\\
The distribution is shown in Figure \ref{cc2cut}.
\noindent
The distribution on the left side in Figure \ref{cl1} is fitted with an exponential function plus a \textit{Gaussian} function, the distribution on the right side of the same figure and the one in the Figure \ref{cc2cut} are fitted with a \textit{Landau} function.\\ 
This analysis is performed for all particles over the whole momentum range; then the fit results are stored and analysed separately to be used during the algorithm implementation.\\
As an example, Figure \ref{meanextra1} shows the distribution of the mean values from the \textit{Gaussian} fit of the kinetic energy for clusters with cluster size equal to 1. The distribution is fitted with an exponential function plus a plateau function.\\
\noindent
Then we focused on the evaluation of the maximum kinetic energy spent to create clusters with cluster size higher than one.\\
To extract this parameter, named \textit{maxExEcl}, we studied the correlation plot, shown in Figure \ref{maxExEcl}, between the total energy loss by particles traversing the gas mixture and the total kinetic energy of clusters with cluster size higher than 1; moreover we evaluated the parameter named \textit{ExSgm} to take into account the smearing around the mean value of the total energy loss.
\noindent
The profile plot is fitted with a linear function and the formula for evaluating the \textit{maxExEcl} is :
\begin{equation}
maxExEcl=\dfrac{E_{tot}-p0+Random(Gaus(0,ExSgm))}{p1}
\end{equation}
where p0 and p1 are the fit parameters of the linear fit and $E_{tot}$ is the total energy loss by the particles traversing the 200 cells of gas.\\
All those elements enter in the algorithm implementation.
\subsection{Three different algorithms} 
\begin{figure}[h]
	\centering
	\includegraphics[width=0.4\textwidth]{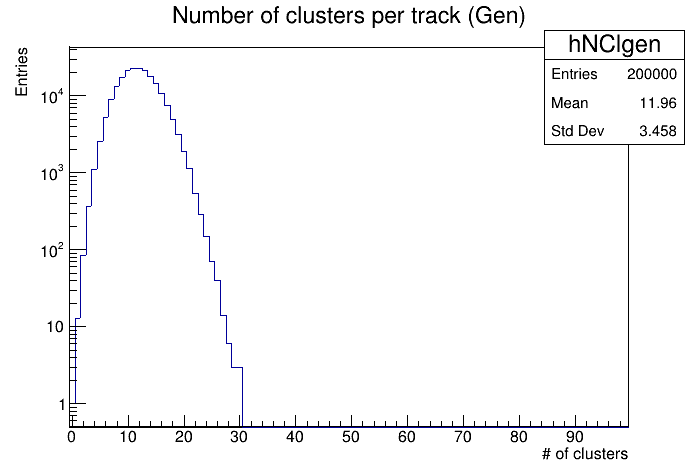}	\includegraphics[width=0.4\textwidth]{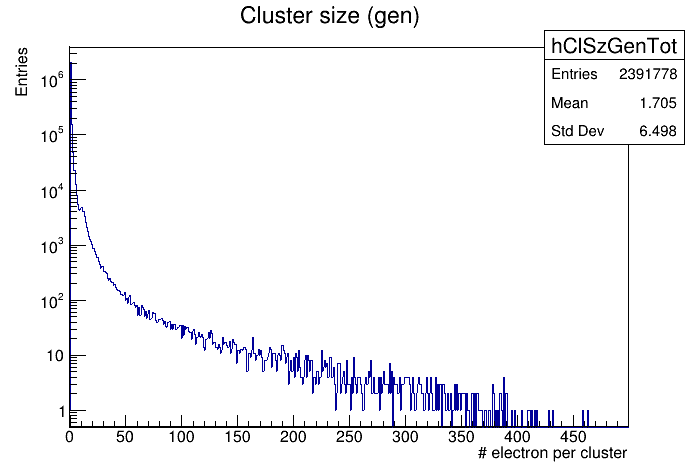}	
	\caption{Clusters number distribution (left) and cluster size distribution (right) for a muon at 300 MeV traversing 200 cells, 1 cm per side, filled with 10 $\% $ He and 90 $\%$ $iC_{4}H_{10}$}
	\label{muon}		
\end{figure}

Using the results from the \textit{Garfield++} analysis described in the section above, we implemented three different algorithms which gave results consistent with the ones expected from simulations.\\
The first algorithm, if the \textit{maxExEcl} is higher than zero, generates the kinetic energy for clusters with cluster size higher than 1 by using its distribution and evaluates the cluster size. This procedure is repeated until the sum of primary ionization energy and the kinetic energy per cluster saturate the \textit{maxExEcl} of the event. Then, using the remaining energy, the algorithm creates clusters with cluster size equal to 1 by assigning their kinetic energy according to the proper distribution.\\
The second algorithm (similar to the previous), during the generation of clusters with cluster size higher than 1, assigns the kinetic energy to them, choosing the best over five extractions that makes the total kinetic energy for cluster with cluster size higher than 1 better approximating the \textit{maxExEcl}.\\
The last algorithm follows a different methodology, since it uses the total kinetic energy of the event to evaluate "a priori" the number of clusters, applying the most likelihood criterium.\\
To verify the validity of the algorithms, we compare their results with the ones simulated by \textit{Garfield++}, taking as case of study a muon with a momentum of 300 MeV, whose clusters number and cluster size distributions are shown in Figure \ref{muon}.
\subsubsection{Results for first algorithm}
\begin{figure}[H]
	\centering
	\includegraphics[width=0.4\textwidth]{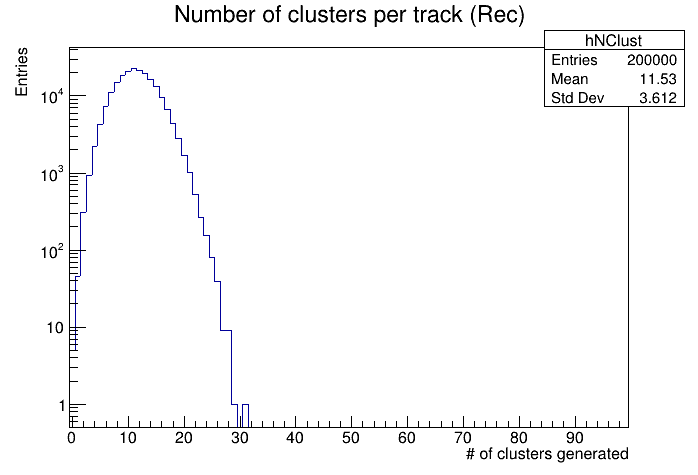}	\includegraphics[width=0.4\textwidth]{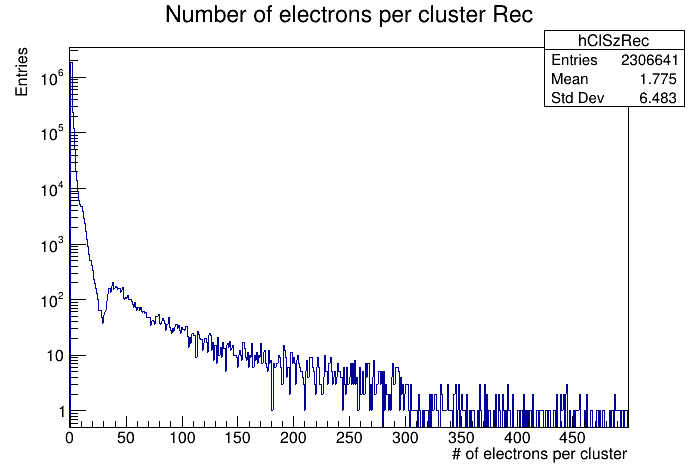}	
	\caption{Clusters number distribution (left) and cluster size distribution (right) for a muon at 300 MeV traversing  200 cells, 1 cm per side, filled with 10$ \% $ He and 90 $\%$ $iC_{4}H_{10}$, as reconstructed by the first algorithm.}
	\label{first}		
\end{figure}

The Figure \ref{first} shows the results obtained from first algorithm.
\noindent
The clusters number distribution reconstructed by the algorithm reproduces the \textit{Poissonian} shape expected. In cluster size distribution a mean value compatible with the one expected exists but its shape presents a small dip before the value of 50 electrons.\\
\subsubsection{Results for second algorithm}
\begin{figure}[H]
	\centering
	\includegraphics[width=0.4\textwidth]{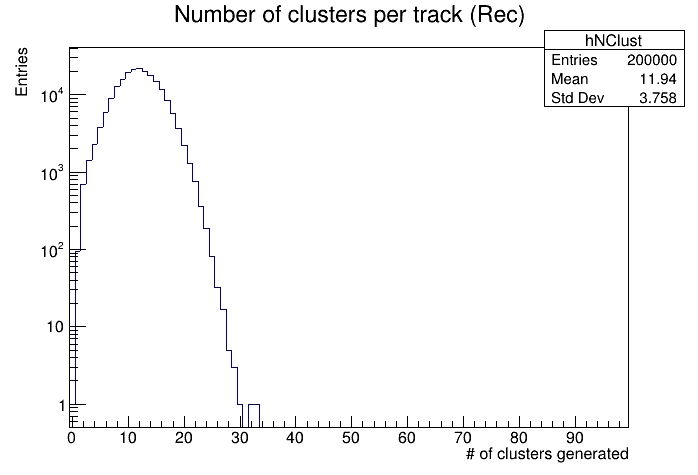}	\includegraphics[width=0.4\textwidth]{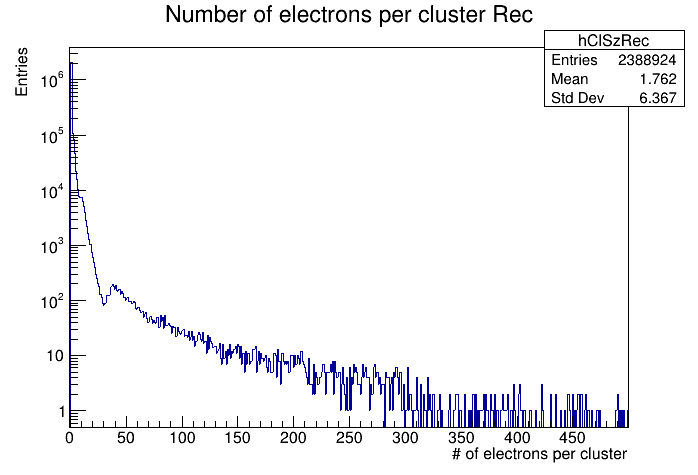}	
	\caption{Clusters number distribution (left) and cluster size distribution (right) for a muon at 300 MeV traversing  200 cells, 1 cm per side,filled with 10$ \% $ He and 90 $\%$ $iC_{4}H_{10}$, as reconstructed by the second algorithm.}
	\label{second}		
\end{figure}

The Figure \ref{second} shows the results obtained from the second algorithm.\\
\noindent
Once again, the cluster number distribution follows the expected \textit{Poissonian} shape. The cluster size distribution gives a mean value consistent with the one expected but the shape has a small dip before 50 electrons.
\subsubsection{Results for third algorithm}
\begin{figure}[H]
	\centering
	\includegraphics[width=0.4\textwidth]{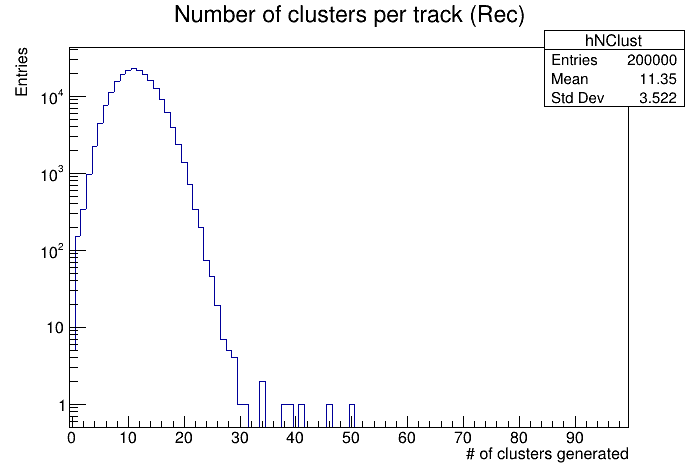}	\includegraphics[width=0.4\textwidth]{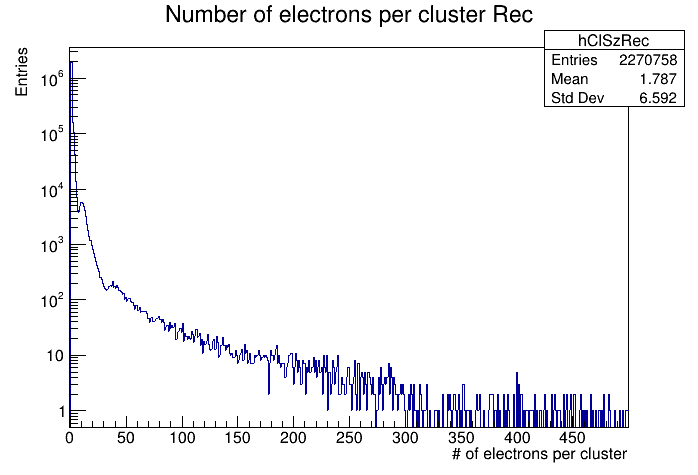}	
	\caption{Clusters number distribution (left) and cluster size distribution (right) for a muon at 300 MeV traversing 200 cells, 1 cm per side, filled with 10$\%$ He and 90$\%$ $iC_{4}H_{10}$, as reconstructed by the third algorithm.}
	\label{third}		
\end{figure}

The Figure \ref{third} shows the results obtained by the third algorithm.\\
\noindent
The third algorithm reproduces the expected clusters number distribution; moreover the cluster size shape is more similar to the one expected than the previous ones obtained from the other two algorithms.\\
The three algorithms are tested also for different particles at different momenta giving consistent results. \\
\section{Cluster counting simulation in Geant4 } 
\begin{figure}[H]
	\centering
	\includegraphics[width=0.5\textwidth]{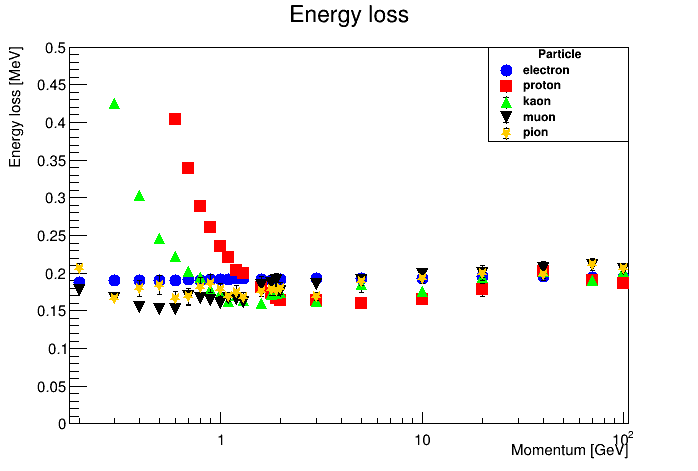}		
	\caption{Energy loss distribution}
	\label{dedxGeant}		
\end{figure}

\begin{figure}[h]
	\centering
	\includegraphics[width=0.4\textwidth]{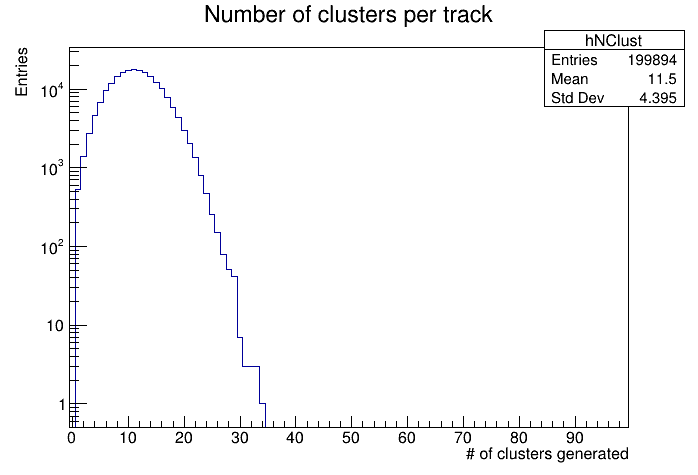}	\includegraphics[width=0.4\textwidth]{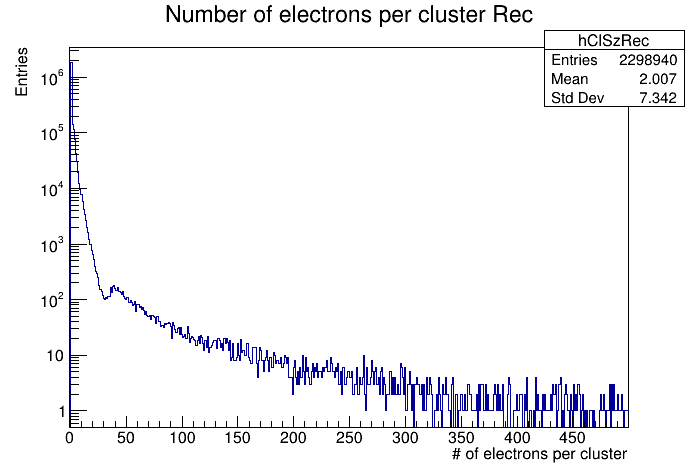}			
	\includegraphics[width=0.4\textwidth]{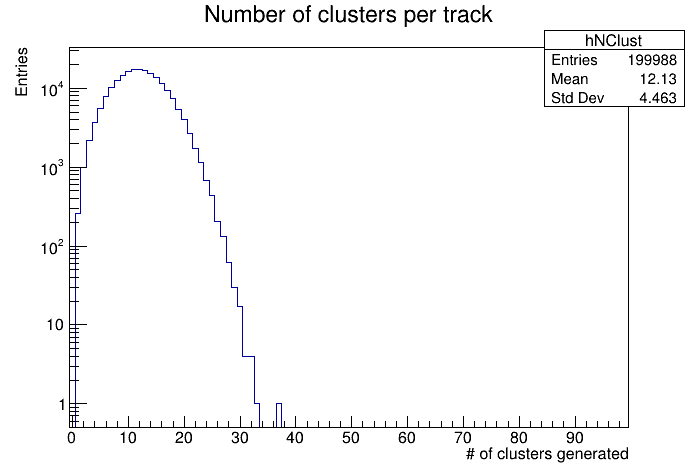}	\includegraphics[width=0.4\textwidth]{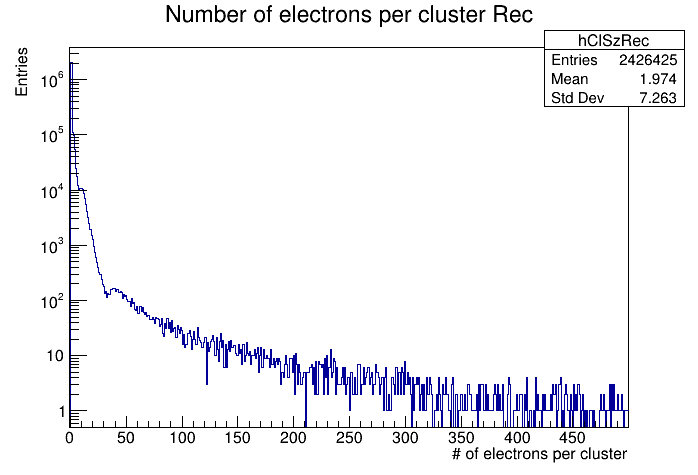}	
	\caption{Clusters number distribution (left) and cluster size distribution (right) for a muon at 300 MeV. On top the ones reconstructed by the first algorithm, on bottom the ones reconstructed by second algorithm.}
	\label{geant4-alg1}		
\end{figure}

\begin{figure}[h]
	\centering
	\includegraphics[width=0.4\textwidth]{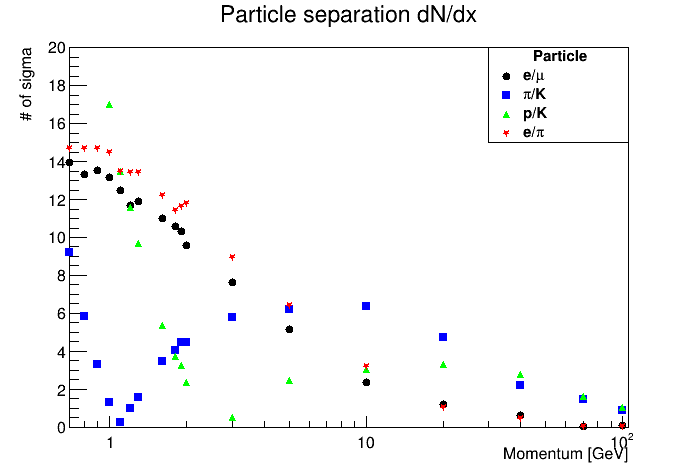}	\includegraphics[width=0.4\textwidth]{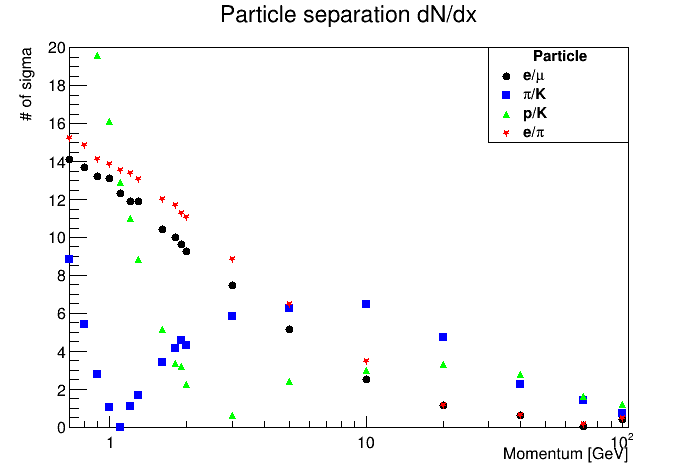}	
	\centering
	\includegraphics[width=0.4\textwidth]{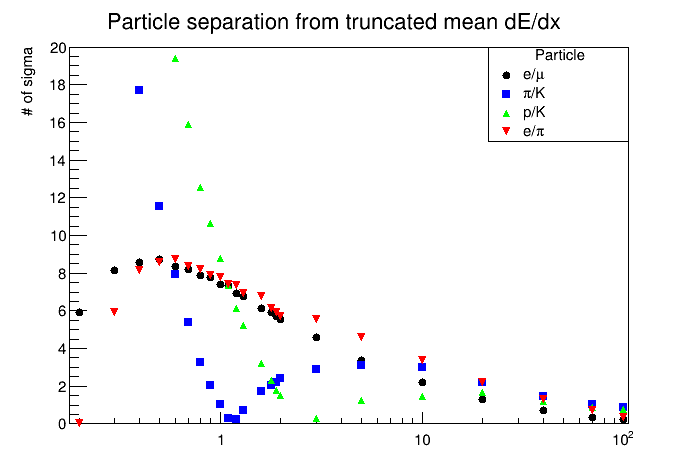}	
	\caption{Top left: The particle separation power obtained with the cluster counting technique implementing the first algorithm. Top right: The particle separation power obtained with the cluster counting technique implementing the second algorithm. Bottom: The particle separation power obtained with traditional dE/dx method.}
	\label{piddedxGeant}		
\end{figure}

The simulations performed in \textit{Geant4} are the same for \textit{Garfield}++, i.e. we simulated 200 cells, 1 cm per side, filled with 10$\%$ He and 90 $\%$ $iC_{4}H_{10}$ traversed by the same five particles in the same momenta range. The energy loss distribution for each particle reconstructed by Geant4 is shown in Figure \ref{dedxGeant}.\\
\noindent
The energy loss is the only necessary information to be provided to the algorithms, whose results (for the first and the second one), are shown respectively on top and bottom of Figure \ref{geant4-alg1}.\\
The two algorithms give consistent results, so we investigate the particle separation power by applying the cluster counting technique and we compare the results with the particle separation power obtained by implementing the traditional method of dE/dx with a truncated mean of 70 $\%$, as shown in Figure \ref{piddedxGeant}.\\
Once again, the simulations confirm that the cluster counting technique improve the particle separation capabilities compared to the traditional method of dE/dx.
\newpage
\section{Conclusions}
We studied the potentials of the cluster counting technique to prove their validity for particle identification with respect to the traditional method, based on the truncated mean of the energy deposited. \\
We implemented also three algorithms that reproduce the clusters number distribution and the cluster size distribution in \textit{Geant4} for the development of the full simulation of the \textbf{IDEA} drift chamber. \\
The next steps are:
\begin{itemize}
	\item to implement the algorithm in a full simulation of the idea drift chamber,
	\item tp perform particle identification studies with the full detector simulation,
	\item to improve Clustering algorithm validation with measurements.
\end{itemize}



%
%


%
%
%
%
%
%
%
%
\end{document}